\documentstyle{article}
\begin{document}
\renewcommand{\thefootnote}{\fnsymbol{footnote}}
\begin{flushright}
KEK-preprint-93-215\\
TUAT-HEP 94-1\\
DPNU-94-11\\
TIT-HPE-94-01\\
NWU-HEP 94-01\\
OCU-HEP 94-1\\
PU-94-681\\
KOBE-HEP 94-03\\
\end{flushright}
\includegraphics{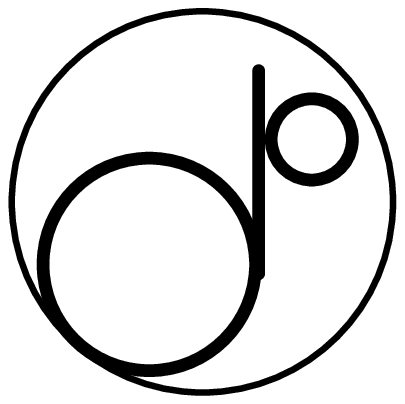}
\begin{flushleft}
{\Large \bf
Measurement of the $D^{*\pm}$ Cross Section
using a Soft-Pion Analysis
in Two-Photon Processes
\footnote{published in Phys. Lett. {\bf B 328} (1994) 535.}
}\\
\vskip 0.5cm
TOPAZ Collaboration\\
\vskip 0.5cm
\underline{R.Enomoto}$^{a,}$\footnote{Email address: enomoto@kekvax.kek.jp},
K.Abe$^{b}$,
T.Abe$^{c}$,
I.Adachi$^{a}$,
M.Aoki$^{c}$,
M.Aoki$^{d}$,
S.Awa$^{e}$
R.Belusevic$^{a}$,
K.Emi$^{b}$,
H.Fujii$^{a}$,
K.Fujii$^{a}$,
T.Fujii$^{f}$,
J.Fujimoto$^{a}$,
K.Fujita$^{g}$,
N.Fujiwara$^{e}$,
H.Hayashii$^{e}$,
B.Howell$^{h}$,
N.Iida$^{a}$,
H.Ikeda$^{a}$,
R.Itoh$^{a}$,
H.Iwasaki$^{a}$,
M.Iwasaki$^{e}$,
R.Kajikawa$^{c}$,
K.Kaneyuki$^{d}$,
S.Kato$^{i}$,
S.Kawabata$^{a}$,
H.Kichimi$^{a}$,
M.Kobayashi$^{a}$,
D.Koltick$^{h}$,
I.Levine$^{h}$,
S.Minami$^{d}$,
K.Miyabayashi$^{c}$,
A.Miyamoto$^{a}$,
K.Muramatsu$^{e}$,
K.Nagai$^{j}$,
T.Nagira$^{e}$,
E.Nakano$^{c}$,
K.Nakabayashi$^{c}$,
O.Nitoh$^{b}$,
S.Noguchi$^{e}$,
F.Ochiai$^{k}$,
Y.Ohnishi$^{c}$,
H.Okuno$^{i}$,
T.Okusawa$^{g}$,
K.Shimozawa$^{c}$,
T.Shinohara$^{b}$,
A.Sugiyama$^{c}$,
N.Sugiyama$^{d}$,
S.Suzuki$^{c}$,
K.Takahashi$^{b}$,
T.Takahashi$^{g}$,
M.Takemoto$^{e}$,
T.Tanimori$^{d}$,
T.Tauchi$^{a}$,
F.Teramae$^{c}$,
Y.Teramoto$^{g}$,
N.Toomi$^{e}$,
T.Toyama$^{c}$,
T.Tsukamoto$^{a}$,
S.Uno$^{a}$,
T.Watanabe$^{d}$,
Y.Watanabe$^{d}$,
A.Yamaguchi$^{e}$,
A.Yamamoto$^{a}$, and
M.Yamauchi$^{a}$\\
\vskip 0.5cm
{\it
$^{a}$National Laboratory for High Energy Physics, KEK,
  Ibaraki-ken 305, Japan
\\
$^{b}$Dept. of Applied Physics,
Tokyo Univ. of Agriculture and Technology,
 Tokyo 184, Japan
\\
$^{c}$Department of Physics, Nagoya University,
 Nagoya 464, Japan
\\
$^{d}$Department of Physics, Tokyo Institute of Technology,
     Tokyo 152, Japan
\\
$^{e}$Department of Physics, Nara Women's University,
 Nara 630, Japan
\\
$^{f}$Department of Physics, University of Tokyo,
   Tokyo 113, Japan
\\
$^{g}$Department of Physics, Osaka City University,
 Osaka 558, Japan
\\
$^{h}$Department of Physics, Purdue University,
 West Lafayette, IN 47907, USA
\\
$^{i}$Institute for Nuclear Study, University of Tokyo,
   Tokyo 188, Japan
\\
$^{j}$The Graduate School of Science and Technology,
Kobe University,
Kobe 657, Japan
\\
$^{k}$Faculty of Liberal Arts, Tezukayama Gakuin University,
 Nara 631, Japan
}

\end{flushleft}

\begin{abstract}
The differential cross section of
$d\sigma(e^+e^-\rightarrow e^+e^-D^{*\pm}X)/dP_T$
was measured using a soft-pion analysis of $D^{*\pm}\rightarrow
\pi_s^\pm D^0(\overline{D^0})$ at TRISTAN.
The average
$\sqrt{s}$ was 58.1 GeV and the integrated luminosity
used in this analysis was
198 pb$^{-1}$, respectively.
\end{abstract}

\section{Introduction}

Multi-hadron production in two-photon processes is qualitatively described
using
the vector-meson dominance model (VDM), the quark-parton model
(direct process) \cite{r1}, and resolved photon processes
\cite{r34,r35,r36,r3,r4,r5}.
In order to understand them quantitatively, charm production is
a good proof, because the theoretical calculations have been
completed to higher
order (order $\alpha_s$) \cite{r16,r23}.
In addition, cut-off parameters, such as $P_T^{min}$, are not
necessary in the theoretical calculation \cite{r6,r22} and the
background from the VDM is considered to be very small.
Studies of charm-meson production should provide useful information
concerning the gluon density in a photon, the current charm quark mass,
and the intrinsic $P_T$ distribution of the
partons in resolved photons.
The statistics of the previous experiments were
too poor to determine these parameters \cite{r7,r8,r9}.


In a previous measurement by TOPAZ using the decay mode
$D^{*+}\rightarrow \pi_s^+ D^0 (D^0 \rightarrow K^-\pi^+ X)$
as well as its
charge-conjugation mode (CC) \cite{r26}, an excess in
charm production compared with the theoretical expectation
was shown \cite{r23}.
Super-symmetric particle production ($\tilde{t}$)
is one possible interpretation of this experimental
excess \cite{r17,r18,r19,r20,r28}.
However, the accuracy of the experiment was
not good enough to justify
this assumption.
For the above-mentioned reasons,
more precise measurements of charm
production in two-photon processes have been awaited.

\section{Analysis}
\subsection{Soft-pion analysis of $D^{*+}\rightarrow \pi_s^+ D^0$}


The decay $D^{*\pm}\rightarrow \pi_s^{\pm}D^0(\bar{D}^0)$ is
characterized by
its small Q value. The maximum transverse momentum
of $\pi_s^\pm$ (soft-pion or $\pi_s$ from
hereafter) with respect to the $D^0$
(from hereafter, the descriptions of the charge states include their
charge conjugations)
flight direction is only 40 MeV.
Its direction can be approximated by the jet-axis obtained by
an invariant mass
algorithm \cite{r27}.
Therefore, from the distribution of the transverse momenta of soft-pions
with respect to the jet-axes ($P_{T,jet}^2$),
the production cross section of
$D^{*\pm}$ can be precisely measured. The acceptance of this
method was proven to be one order higher than
that of the exclusive reconstruction
of $D^{*\pm}$ decay \cite{r27}.
We thus tried this method for two-photon events,
and obtained the differential cross section of $D^{*\pm}$
as a function of $P_T$,
the transverse momentum with respect to the beam axis.
Higher statistics and lower systematic error
measurements were expected using this analysis.

\subsection{Data sample}


The data were obtained by the TOPAZ detector at the TRISTAN
$e^+e^-$ collider of KEK. The details concerning the TOPAZ detector can be
found elsewhere \cite{r10}.
The integrated luminosity of the event sample used in the analysis
was 198 pb$^{-1}$. The average $\sqrt{s}$ was 58.1 GeV.
The trigger conditions were as follows:
more than two tracks with $P_T$ $>$ 0.3 $\sim$ 0.7 GeV
and an opening angle $>$ 45 $\sim$ 90 degrees (depending on the beam
condition); the neutral energy deposit in the
barrel calorimeter be greater than 2 $\sim$ 4 GeV; or that in the
endcap calorimeter be greater than 10 GeV.

The event selection criteria for the two-photon process were tighter
than in the exclusive $D^{*\pm}$ analysis \cite{r26}, because this
signal suffered from the background of the VDM as well as
single-photon exchange hadronic events.
The selections were as follows: the number of charged tracks be
$\geq$ 4; the total visible energy in the central part of the detector
be between 5 and 25 GeV; the vector sum of the transverse momenta of
the particles with respect to the beam axis be less than 7.5 GeV;
the absolute value of the sum of charges be $\leq$ 3;
the absolute value of the cosine of the thrust axis in the laboratory
frame be less than 0.9, where neutral clusters detected by the endcap
calorimeters were included;
and no large energy clusters in the
barrel calorimeter ($E>0.25E_{beam}$).
In addition, we divided each event into two jets with respect to the
plane perpendicular to the thrust axis at the laboratory frame;
the cosine of the
angle between two jets was required to be greater than -0.9.
These cuts were determined using a Monte-Carlo simulation
of the direct process events so as to maximize the charm
event acceptance and its purity.
A total of 14128 events were selected.

\subsection{Particle selection}


The charged track selections were as follows: the closest approach
to the event vertex be consistent within the measurement error;
the number of degrees of freedom (DOF) in the track-fitting
be $\geq$ 3; and $P_T$ be $\geq$ 0.1 GeV.
The $\gamma$ selections were: the cluster be detected by a barrel
calorimeter; the energy be $\geq$ 0.2 GeV; and the cluster position be
separated from any charged-track extrapolations
by more than 10 cm at the surface of the barrel calorimeter.

\subsection{Jet-axis determination}


The jet-axis was obtained by using an invariant mass
algorithm \cite{r27}, in which particles were merged
together in an iterative way if their invariant mass was less than
2 GeV. In this algorithm, a soft-pion candidate, whose
$P_{T,jet}^2$ must be calculated, was removed from the sample
particles, i.e., the jet-axis was recalculated each time when
a particle was selected to be the soft-pion candidate.
Then, those jet candidates having invariant
masses greater than 0.3 GeV and transverse momenta with respect
to the beam-axis of greater than 1 GeV were selected.
The former cut were to remove jets made of single track.
The fake tracks by the patter recognition
(duplicating tracks using the same space-points)
were removed from the
charged track sample.
By the above algorithm, two or three jets were usually
reconstructed (85\% of the events).

Those cut values were determined using a Monte-Carlo simulation
while assuming a direct process. The best angular resolution of
the soft-pions with respect to the jet-axes
was obtained using the above set of cut values.
Even when the jet-mass was as low as 0.3 GeV, the $D^0$ direction
was calculated with the acceptable resolution which was proven
by a Monte-Carlo simulation.
The
estimated angular resolution of the soft-pions was
approximated by $e^{3.7-4.3P_T}$ degrees ($P_T$ in GeV).
The transverse momentum resolutions with respect to the jet-axis
were $70\pm10$ MeV in the $P_T$ region between 0.1 and 0.6 GeV,
a slowly increasing function of the soft-pion's $P_T$.

The transverse momentum of the soft-pion was calculated with respect
to the closest jet candidate ($P_{T,jet}^2$).

The effects by the pair conversion electron and Dalitz pairs
were estimated using a Monte-Carlo simulation.
They were proven to be absorbed in the background function in the
fitting procedure which is described in Section \ref{sec262}.

\subsection{$P_{T,jet}^2$ distribution of soft-pions}

\begin{figure}
\vspace{15cm}
\includegraphics{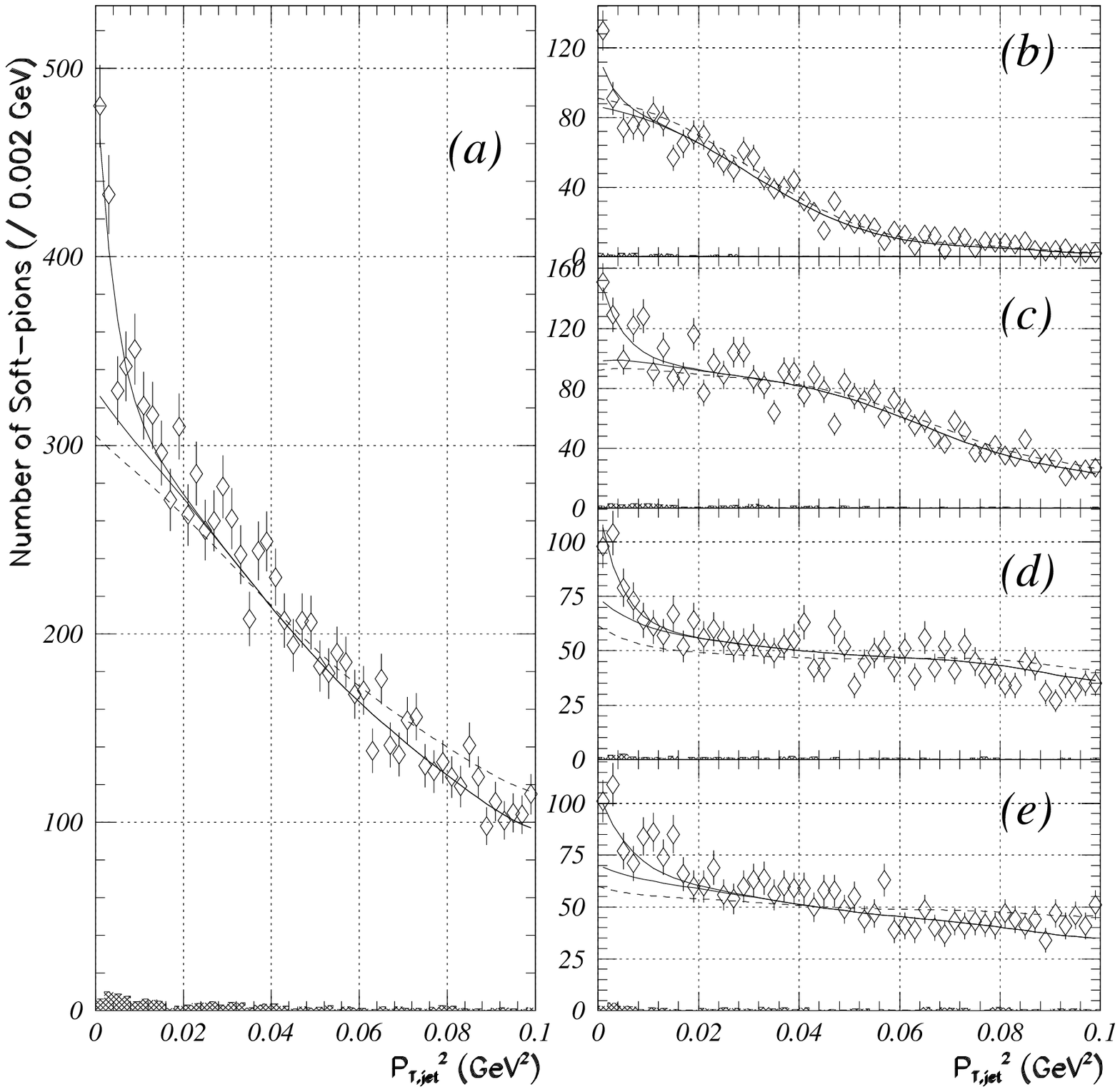}
\caption{
Distributions of $P_{T,jet}^2$ of soft-pion candidates:
(a) for $0.1<P_T<0.6$ GeV, (b) $0.1<P_T<0.2$, (c) $0.2<P_T<0.3$,
(d) $0.3<P_T<0.4$, and (e) $0.4<P_T<0.6$.
The dashed lines are the result of a Monte-Carlo simulation
of the background, where the normalizations were carried out using the
entries of $P_{T,jet}^2>0.02$ GeV.
The hatched histograms are contributions from
$e^+e^-\rightarrow (\gamma)\rightarrow D^{*\pm}X$.
The solid curves are the best-fitted functions
which are
described in the text.
The higher lines are signals and backgrounds,
and the lower are backgrounds.
}
\label{fig1}
\end{figure}

The $P_T$ range of the soft-pion candidates was set to
$0.1<P_T<0.6$ GeV in order to cover the predicted region
by the direct process.
The $P_{T,jet}^2$ distribution for the above-mentioned $P_T$ range
is plotted in Figure \ref{fig1} (a).
There is a clear peak at around the zero $P_{T,jet}^2$ region.

In order to derive the differential cross section with respect to $P_T$,
we selected the following $P_T$ ranges: (b) $0.1<P_T<0.2$, (c)
$0.2<P_T<0.3$, (d) $0.3<P_T<0.4$, and (e) $0.4<P_T<0.6$ GeV.
The corresponding distributions are shown in Figures \ref{fig1}
(b), (c), (d), and (e), respectively.
The binning was selected so as to give similar peak entries.
The resolution of the $P_T$'s of the soft-pions are smaller than these
binnings.

\subsection{Background estimation}

\subsubsection{Monte-Carlo simulations}

In the soft-pion analysis, background emulation using the
experimental data is difficult,
in contrast to the case
in a full-reconstruction analysis, where
wrong sign combinations
can be used to emulate the background \cite{r26}.
We therefore used a Monte-Carlo method.
The $\gamma\gamma$ generation was based on an equivalent photon
approximation. For the parton generations,
we used lowest-order formulas (Born approximation).
In the resolved photon processes, LAC1 \cite{r4}
parametrization was used for the parton density in the photon.
For light-quark event generation, a $P_T^{min}$ of 2.5 GeV was
used, which was the best-fitted value using the general
event shapes of a $\gamma\gamma\rightarrow$ (multi-hadrons) obtained by
the TOPAZ detector \cite{r22}. In addition, the VDM Monte-Carlo
events and $e^+e^-\rightarrow \gamma \rightarrow q\bar{q}$
events by the LUND 6.3 (parton shower option) program were
added. In LUND 6.3, the branching ratios of $D^{0}$
were adjusted to the PDG values \cite{r13}.
Hadronization of partons generated by the $\gamma \gamma$ events
was carried out using LUND 6.3 string fragmentation without a
parton shower option \cite{r12}.
A more detailed description
together with a detector simulation
can be found in Reference \cite{r22}.

The branching ratio of $D^{*+}\rightarrow \pi_s^+ D^0$ was
assumed to be 68.1\%, which was obtained by CLEO \cite{r14}.
The V/(V+P) ratio was set to 0.75 and a u:d:s:qq ratio of 1:1:0.3:0.1
was used.

The results of this Monte-Carlo simulation are shown in Figures
\ref{fig1} (a) - (e) by the dashed lines. The
events with $D^{*\pm}\rightarrow \pi^\pm D^0(\overline{D^0})$
were removed from these.
They fit with the experimental data in the higher
$P_T$ regions.
In the lower $P_T$ region, however, since the agreement of the normalization
scale was poor, the lines in Figure \ref{fig1} were normalized
using the entries of $P_{T,jet}^2>0.02$ GeV.
However, the background shape (i.e., in the higher $P_{T,jet}^2$ region)
agreed well.
The contributions from $e^+e^-\rightarrow \gamma \rightarrow
D^{*\pm}X$ are shown in Figures \ref{fig1} (a) - (e) by
hatched histograms, and are considered to be negligible.

\subsubsection{Fit of the $P_{T,jet}^2$ spectra}
\label{sec262}

There is a difference between the experimental and Monte-Carlo
data in the lower $P_T$ region, especially for the normalization scale.
We conclude that this is due to the facts that
the $c\bar{c}$ cross section in the Monte-Carlo
program was not correct
(also, $P_T^{min}$ for light quark events was not correct)
and that the higher-order effects (order $\alpha_s$)
were not included (for both $c\bar{c}$ and light quark-pair events).
The small differences appeared in the slope in the $P_{T,jet}^2$ spectrum.
We therefore multiplied
the smoothed
Monte-Carlo background by
a first-order polynomial (a+b$P_{T,jet}^2$).
The coefficients (a and b) of the polynomial were
set to be free parameters. The signal shapes
were assumed to be exponential functions $\propto e^{-P_T/\beta}$.
The $\beta$ at each binning was obtained by fitting
Monte-Carlo signal events, and was used as a fixed parameter in the
fitting procedures, because the soft-pion angular resolution seemed to be
consistent with the prediction of the detector simulation.

The results of the signal and
background estimations are shown in Figures \ref{fig1}
(a) - (e) by the solid lines. The peak entries obtained by the
above-mentioned
procedures are $50\pm32$, $126\pm26$, $83\pm20$, and $113\pm29$,
in the 0.1 - 0.2, 0.2 - 0.3, 0.3 - 0.4, and 0.4 - 0.6 GeV
$P_T$ regions,
respectively. In total, we obtained $372\pm54$ events.

The main systematic error of the analysis is due to this fitting
procedure, since we did not know the exact shapes of the
background. We tried various background
shapes, such as $(1+aP_{T,jet}^2+bP_{T,jet}^4+cP_{T,jet}^6)^{-1}$ by ALEPH
\cite{r15} and
$(1+(P_{T,jet}^2/a)^b)^{-1}$ by TOPAZ \cite{r27}.
We defined the systematic errors of the fitting procedures to be the
differences in the obtained peak entries by the various background
functions. In addition, we tried to simulate various sets of the
Monte-Carlo events by changing the generation parameters
described in the previous subsection, and carried out the same
fitting procedures. By these procedures, we obtained systematic errors
of 12, 23, 18, and 9\% in the 0.1 - 0.2,
0.2 - 0.3, 0.3 - 0.4, and 0.4 - 0.6 GeV $P_T$ ranges, respectively.

\subsection{Experimental checks of the data sample}

\begin{figure}
\vspace{15cm}
\includegraphics{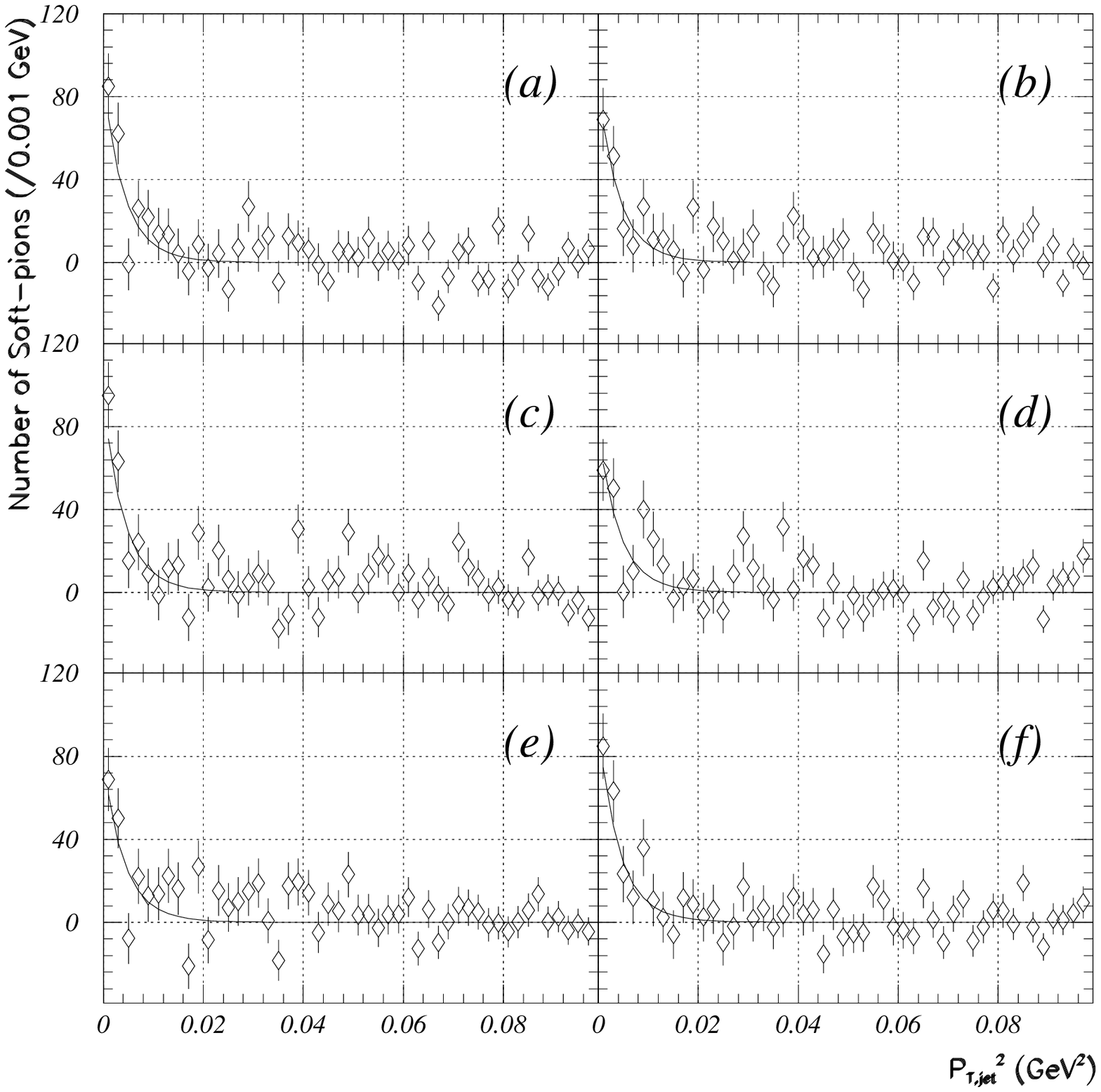}
\caption{
Distributions of $P_{T,jet}^2$:
(a) for $\pi_s^+$, (b) $\pi_s^-$, (c) in the forward direction,
(d) in the backward
direction, (e) for positive charge asymmetry, and (f) for negative
charge asymmetry.
The data points are
background-subtracted signals using the fitting procedure
described in the text.
The solid curves are the best-best fitted signal lines.
}
\label{fig2}
\end{figure}

We carried out some miscellaneous checks on these peaks.
In order to check
detector bias, the charge states
and the cos$\theta _{\pi _s}$ distribution were tested.
The background-subtracted $P_{T,jet}^2$ distributions
are shown in Figures \ref{fig2} (a) - (d):
(a) for $\pi_s^+$, and (b) $\pi_s^-$, (c) cos$\theta_s>0$,
and (d) cos$\theta_s<0$.
If $D^{*\pm}$'s had originated from single-photon exchange events
(for example cascade decays of $b\bar{b}$ events which were
produced at low angles with respect to the beam axis),
some asymmetries may appear in the charge states,
or the cos$\theta _{\pi _s}$ distribution.
These are shown in Figures \ref{fig2} (e) and (f):
(e) charge of $\pi_s$ multiplied by cos$\theta_s$
be $>0$, and (f) that be $<0$.
The pairs
(a) and (b), (c) and (d), and (e) and (f) seem to be consistent with
each other.
The numbers of signals in these plots are 190, 182, 202, 169,
168, and 203, respectively, where the statistical errors of them
are about 34.
In the case of single-photon exchange events ($b\bar{b}$ cascade),
the asymmetry of 1:2 should be observed between Figure \ref{fig2}
(e) and (f).
In addition, the same checks on the higher $P_T$ events
($0.3 < P_T < 0.6$ GeV) were carried out. The peak entries
corresponding to the definitions (a)-(f) were
97, 99, 89, 107, 102, and 94 with the statistical errors of
about 20, respectively.
They are also consistent with each other.
We thus conclude that these events did not originate from
single-photon processes.

\section{Production cross section}

\subsection{Systematic ambiguities}

This analysis was based on momentum measurements of
inclusive pions and jet-axis reconstructions.
The former part was well defined and the systematic errors
were well known from a previous physics analysis using the
TOPAZ detector \cite{r29}.
The largest systematic error in the tracking algorithm appears
especially in the low-$P_T$ region, i.e., the 0.1-0.2 GeV region.
The systematic errors were estimated using hadronic events via
single-photon exchange processes and the LUND 6.3 Monte-Carlo
simulation, since this Monte-Carlo method was known to reproduce
many experiments.
The thus-obtained systematic errors due to tracking were 7
and 3\% at the 0.1 - 0.2 and 0.2 - 0.6 GeV $P_T$ regions,
respectively.

The systematic error due to the jet-axis determination were evaluated
by changing the maximum value of the invariant mass and the
transverse momentum cut in selecting jets by 10\%.
The errors were 11, 10, 6, and 3\% in the
0.1 - 0.2, 0.2 - 0.3, 0.3 - 0.4, and 0.4 - 0.6 GeV
$P_T$ ranges, respectively.

The systematic errors in the event selection
were estimated
in the same way by changing the cut values, such as
the visible energy
and missing $P_T$, by 10\%, and the sum of charges by one.
Those due to the hardware triggers were estimated using a trigger
simulation program. For a charged trigger, we added 5\% of accidental hits
in the tracking detectors. The acceptance was increased by $2\pm1.4$\% by this.
The percentage of neutral triggered events was only 5.6\%. Therefore
the systematic errors caused by the energy calibrations of calorimeters
and the summing amplifiers noises were considered to be negligible
small.
In total, the systematic errors were estimated to be 7\%.

In total, the systematic errors in determining the
production cross sections of the soft-pions were
19, 26, 20, and 12\% in the
0.1 - 0.2, 0.2 - 0.3, 0.3 - 0.4, and 0.4 - 0.6 GeV
$P_T$ ranges, respectively.
{}From now on, the errors include both the statistical
and systematic contributions.

\subsection{Soft-pion cross section}

\begin{figure}
\vspace{7cm}
\includegraphics{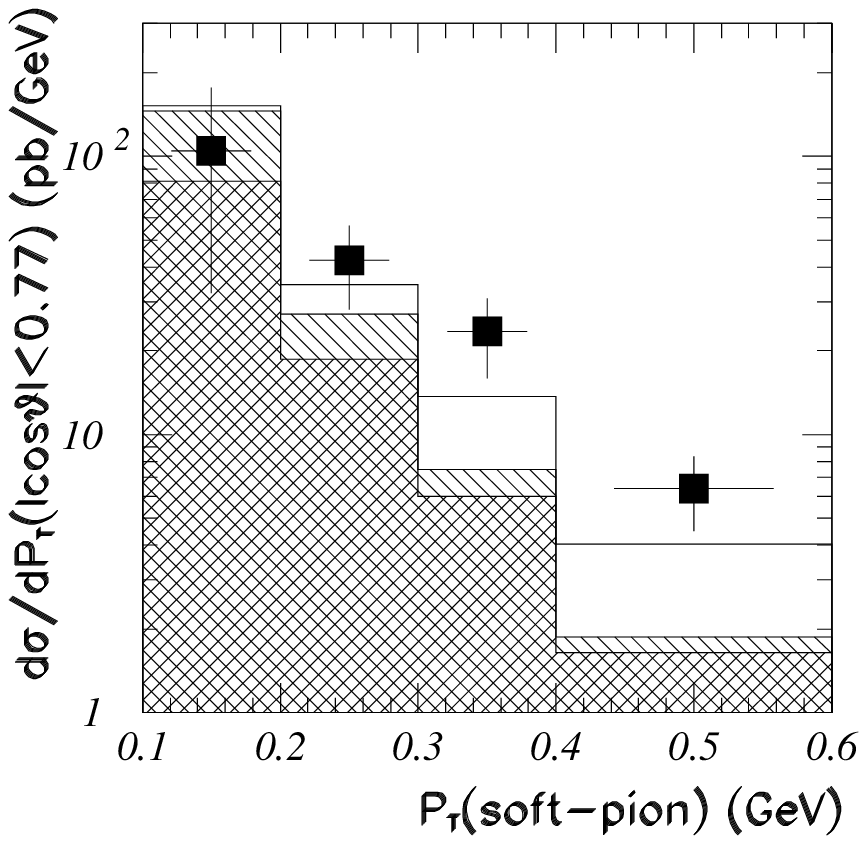}
\caption{
Differential cross section of the soft-pions versus
$P_T$.
The histograms are the
theoretical predictions: the cross-hatched area is the direct process,
the singly-hatched area the resolved process
(LAC1), and the open area $\tilde{t}$ pair production.}
\label{fig3}
\end{figure}

We obtained the $P_T$ differential cross section for soft-pions.
The $P_T$ of a soft-pion is almost proportional to that of the $D^{*\pm}$,
and also to that of the charm quark.
In order to define the detector's sensitive area,
we restricted the cosine of the soft-pion emission angle
with respect to the beam axis to be within $\pm$0.77.
Then, the acceptance differences, especially between the direct and
resolved photon processes, became small (
+5\% higher for the resolved photon process).
The acceptance was obtained using a Monte-Carlo
simulation of the direct and resolved (LAC1) photon processes
and it was a increasing function of $P_T$.
The thus-obtained differential cross section is shown in Figure \ref{fig3}
and is listed in Table \ref{table1}, together with the theoretical
predictions (described later).
\begin{table}
\begin{center}
\begin{tabular}{ccccc}
\hline
\hline
$P_T$ range (GeV) & Experiment
& Direct & Direct+LAC1 & Direct+LAC1+$\tilde{t}\bar{\tilde{t}}$ \\
\hline
0.1 - 0.2 & $105\pm72$ & 81 & 145 & 152 \\
0.2 - 0.3 & $42.3\pm14.2$ & 18.7 & 27.1 & 34.6 \\
0.3 - 0.4 & $23.4\pm7.5$ & 6.0 & 7.5 & 13.6 \\
0.4 - 0.6 & $6.4\pm1.9$ & 1.6 & 1.9 & 4.0 \\
\hline
\hline
\end{tabular}
\end{center}
\caption{$d\sigma/dP_T(soft-pion)(|cos\theta|\leq 0.77)$ (pb/GeV).}
\label{table1}
\end{table}

\subsection{$D^{*\pm}$ production cross section}

\begin{figure}
\vspace{7cm}
\includegraphics{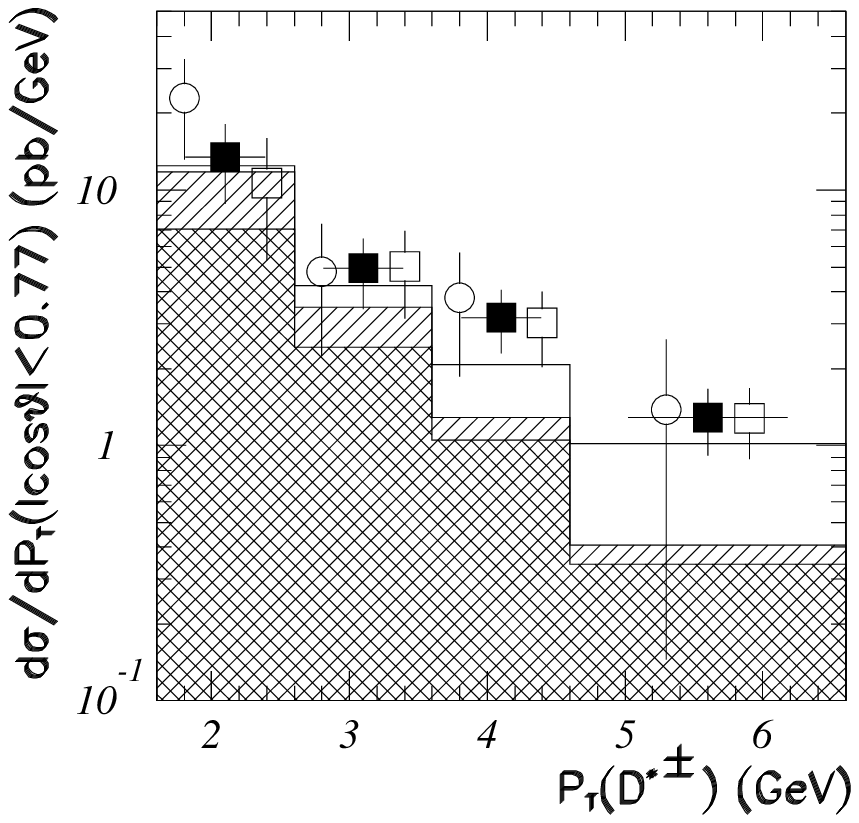}
\caption{
Differential cross section of $D^{*\pm}$ versus
$P_T$. The open circles were obtained from a
previous experiment,
the open squares are from this experiment, and the closed
squares are the combined values. The histogram definitions are
the same as those given in Figure 3.
}
\label{fig4}
\end{figure}

In order to compare this data with
the previous results of the $D^{*\pm}$
production at TRISTAN \cite{r26}, we unfolded the
soft-pion cross section to that of $D^{*\pm}$.
The method was as follows:
\begin{enumerate}
\item We tuned the theory by iteratively changing
parameters such as the charm-quark mass
(current mass) so as to fit the experimental cross section
($d\sigma(D^{*\pm})/dP_T$) as
well as possible. The obtained current charm mass was 1.3 GeV.
The details are described in Section \ref{sec4}.
\item Using the above parameter, we carried out a simulation
and made a conversion matrix from the soft-pion $P_T$
to that of $D^{*\pm}$.
\item We multiplied this matrix by the experimental cross section
of the soft-pions.
\end{enumerate}
The variation in the
$P_T$ spectrum of $D^{*\pm}$ in the Monte-Carlo
was a source of systematic error.
We therefore assumed various $P_T$ spectra of $D^{*\pm}$
by changing the fragmentation function parameters in LUND 6.3 and
tried unfolding.
The systematic error of the unfolding was obtained to be 7\%.
Also there were strong correlations between the neighboring
$P_T$ binnings. The largest one was 85\% between
the binnings 0.1-0.2 and 0.2-0.3 GeV. The magnitudes
of the other binning pairs were 30 $\sim$ 40\%.
The errors due to these correlations were taken into account.
The obtained differential cross section of $D^{*\pm}$
is indicated in Figure \ref{fig4} by the open squares.
The previous results are indicated by the open circles \cite{r26}.
Both experiments are consistent within their errors;
the overlapping statistics of the present analysis are obvious.
The experimental averages of both measurements are also
indicated in Figure \ref{fig4} by the solid squares and are listed
in Table \ref{table2}, along with the theoretical predictions.
\begin{table}
\begin{center}
\begin{tabular}{ccccccc}
\hline
\hline
$P_T$ range & Full & Soft-  &Average
& Direct & Direct+LAC1 & Direct+LAC1+$\tilde{t}\bar{\tilde{t}}$ \\
(GeV) & reconstruction & pion & & & & \\
\hline
1.6 - 2.6 & $22.9\pm9.7$ & $10.7\pm5.3$ & $13.5\pm4.7$
& 7.1 & 11.8 & 12.5 \\
2.6 - 3.6 & $4.80\pm2.56$ & $5.03\pm1.88$ & $4.95\pm1.51$
& 2.42  & 3.48 & 4.22 \\
3.6 - 4.6 & $3.79\pm1.92$ & $3.01\pm0.99$ & $3.17\pm0.88$
& 1.05 & 1.29 & 2.08 \\
4.6 - 6.6 & $1.38\pm1.23$ & $1.28\pm0.40$ &$1.29\pm0.38$
& 0.34 & 0.41 & 1.02 \\
\hline
\hline
\end{tabular}
\end{center}
\caption{$d\sigma/dP_T(D^{*\pm})(|cos\theta|\leq 0.77)$ (pb/GeV).}
\label{table2}
\end{table}

\section{Discussion}
\label{sec4}

\subsection{Predictions of a two-photon process}

\subsubsection{Lowest-order calculation}

The experimental cross section of $D^{*\pm}$
in the region $1.6\leq P_T \leq 6.6$ GeV and $|cos\theta |\leq 0.77$
was $24.2 \pm 5.0$ pb (Table \ref{table2}).
We first compare this with the lowest-order calculation.
A current-charm quark mass of 1.5 GeV and LAC1
parametrization with $\Lambda _{\overline{MS}}$ = 0.2 GeV
were used.
The predictions of the direct and resolved photon processes
were 8.5 and 2.5 pb (11.0 pb in total), respectively,
significantly lower than the experimental observation.

\subsubsection{Higher-order corrections}

Corrections on the order of $\alpha_s$ in the QCD part of the theory
are available in the analytic calculations \cite{r16,r23},
but not in the Monte-Carlo calculations.
Our Monte-Carlo simulation was
based on a lowest-order calculation (LO), which
was followed by string fragmentation using LUND 6.3.
Since string fragmentation includes a parton-shower-like
effect, the next-to-leading-order effect in the
$P_T$ spectrum of $D^{*\pm}$ is already counted.
For single-photon exchange events, the difference in the
momentum spectra of $D^{*\pm}$ between the LO-matrix
element (i.e. two-jet only) and the parton-shower options in
Lund 6.3 is only 9\% at $\sqrt{s}=10$ GeV.
We therefore concluded that the systematic error due to
the hadronization process is 9\%.
We then carried out a next-to-leading-order correction (NLO)
($P_T$-independently) of the direct process. The factor was 1.31,
as described in reference \cite{r23}.
For the resolved photon process,
due to the presence of the process $\gamma q \rightarrow
c\bar{c} q$ (a part of this is absorbed in the gluon density
function in the resolved photon), the $P_T$
dependent factors were obtained in the following way \cite{r24}:
\begin{enumerate}
\item We derived the $P_T$-dependent ratios between the higher-
and lowest-order calculations for both the direct and resolved
photon processes.
\item We then calculated the ratios between these ratios of the
direct and resolved photon processes.
\item The final ratios were normalized so as to fit the total
cross section of the higher-order calculations for the resolved photon
process.
\end{enumerate}
The obtained $P_T^c$-dependent correction were written as
$$0.50P_T^c+0.54,$$
where $P_T^c$ was the $P_T$ of the charm quark in GeV. Here,
the charm quark mass ($m_c$) and the renormalization scale ($\mu$)
of 1.6 GeV and $\mu=\sqrt{2} m_c$ were used, respectively.
This correction factor was used as event weight in the Monte-Carlo
simulations.

Then, the prediction of the cross section in the range
($1.6\leq P_T \leq 6.6$, $|cos\theta | \leq 0.77$)
was 15.6 pb, still lower than the experimental observation.

\subsubsection{Dependence on the current charm quark mass and gluon
$P_T$ distribution}

If the current charm-quark mass is significantly lower than
1.5 GeV, the total cross section of $D^{*\pm}$
becomes large.
For example, the Monte-Carlo package PYTHIA 5.6 used
$m_c$ = 1.35 GeV \cite{r30}.
However, the cross section increased only at a low-$P_T$ region
i.e., $\leq 2.6$ GeV.
The PYTHIA 5.6 program also included a gluon $P_T$ of
0.44 GeV (Gaussian) inside the resolved photon jet.
This may increase the high $P_T$ cross section of $D^{*\pm}$.
We thus lowered $m_c$ to 1.3 GeV and added a Gaussian
fluctuation in the gluon $P_T$ distribution inside
a photon of 0.44 GeV.
The obtained cross section was 17.4 pb.

The differential cross sections which this model predicted
are given in Figures \ref{fig3}, \ref{fig4} and Tables
\ref{table1}, \ref{table2}.
The predictions agree with the experimental data
in the 1.6 - 3.6 GeV $P_T$ region. However, there is a large excess
in the higher $P_T$ regions ($>3.6$ GeV).

When we tried other resolved photon parameterization,
such as DG \cite{r3}, the predicted cross sections
were significantly lower than the experimental data,
even upon changing some of the parameters described so far.

\subsubsection{Ambiguities in the prediction}


We first estimate the ambiguities in the NLO correction.
As has been described, string fragmentation describes a
part of the NLO corrections. Therefore, the ambiguities lie in
the hard process descriptions, such as three jets and
hard bremsstrahlung from the resolved quark line. These are estimated
by using the single-photon exchange events generated by the LUND 6.3
Monte-Carlo program.
The $D^{*\pm}$ momentum spectra between the LO-matrix with
string fragmentation and parton shower options have been compared
at $\sqrt{s}$=10 GeV.
A difference of 9\% was observed; this was considered to be a
systematic ambiguity.

Second, we considered the ambiguities in
the charm mass, renormalization scale ($\mu$), and gluon
intrinsic $P_T$ inside a photon. The ranges that we selected were
$m_c=1.3\pm 0.15$ GeV, $m_c < \mu < 2m_c$,
and $0 < P_T^{gluon} < 0.44$ GeV, respectively.
In total, the cross-section ambiguity was obtained to be 11\%
for the lower $P_T$ region and 9\% for the higher $P_T$ region.

Third, the ambiguity in the equivalent photon approximation
was studied. In our event generation, the initial photon
from beam was considered to be almost real photon with the
direction parallel to the beam axis. We here used a matrix element
calculation by Kuroda \cite{r33} in the event generation of
the direct process. The differential cross section at the higher
$P_T$ region was increased by 25\% by this effect.
We considered that this value is a systematic error due to the
$Q^2$ dependence of $\gamma \gamma$ system in the direct process.
In the resolved photon case, we do not know how to take care of this
effect. We thus approximated that all photons are real photons.
We considered this systematic error
due to the motion of $\gamma \gamma$ system is absorbed in
that of the intrinsic gluon $P_T$ inside the photon.

Last, the threshold effect of the $c\bar{c}$ productions are commented.
The Wvis distribution, i.e., the invariant mass of visible hadronic
system, are shown in Figure \ref{fig6} (a) and (b) for the lower
and higher $P_T$ soft-pion signals. They were dominated in the region
where Wvis $>$ 5 GeV.
The real W$_{\gamma \gamma}$
was expected to be distributed
above this value. We therefore concluded that the threshold effect
was small in our detection region. We did not consider about the systematic
ambiguity due to the threshold enhancement by this reason.

The experimental excess at the high-$P_T$ region
($P_T(D^{*\pm}) >3.6$ GeV) with respect to the theoretical
prediction of the direct plus resolved (LAC1) photon processes
becomes a $2.9 \sigma$ effect.

\subsection{Possibility of $\tilde{t}$ pair production}

\begin{figure}
\vspace{15cm}
\includegraphics{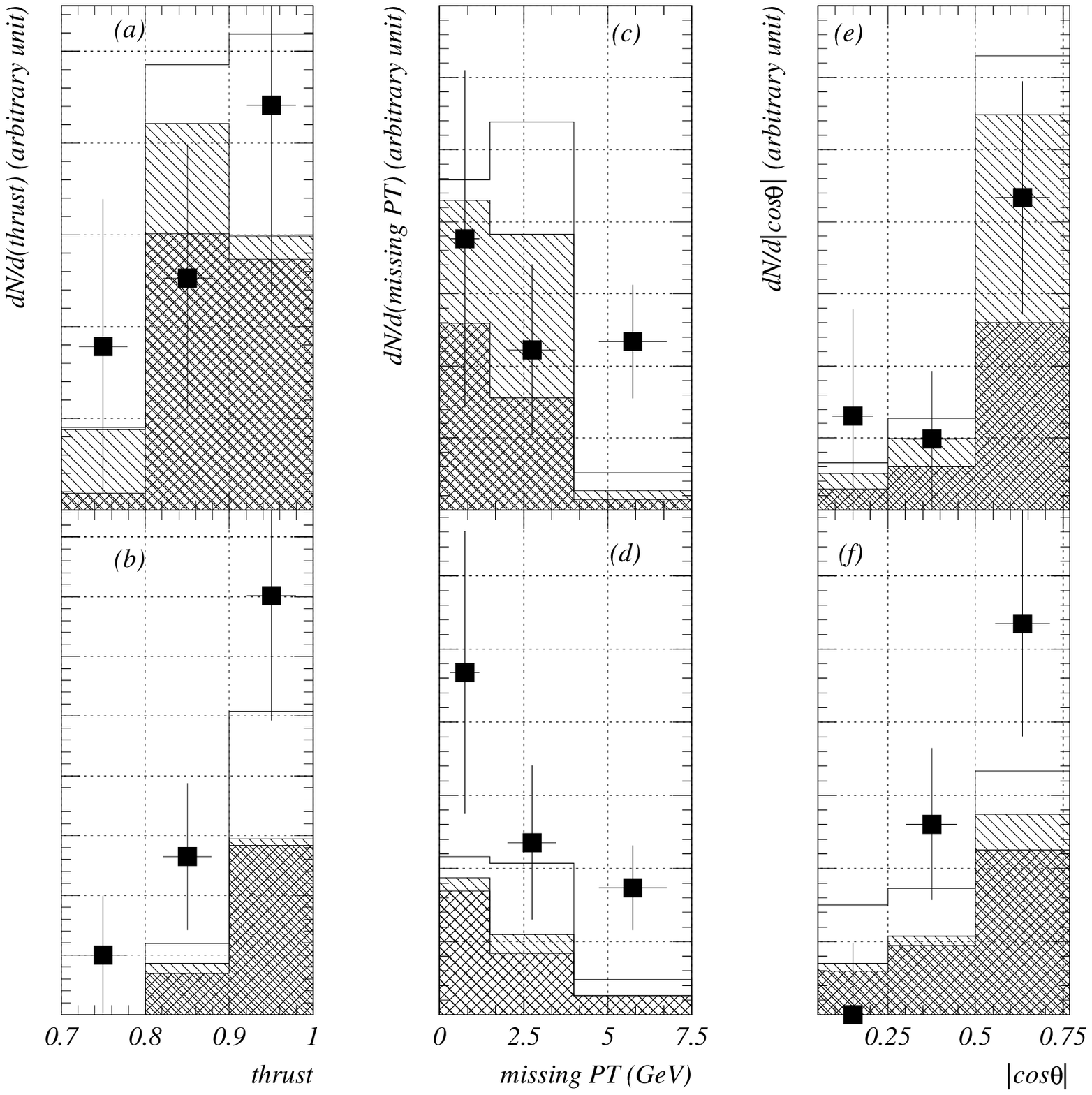}
\caption{
Various event shape distributions for the lower ($P_T<0.3$ GeV)
and higher ($P_T>0.3$ GeV)
$P_T$ events:
(a) thrust distribution in the lower $P_T$ events, (b) that in the higher
$P_T$ region, (c) missing $P_T$ distribution in the lower $P_T$ region,
(d) that in the higher $P_T$ region, (e) $|cos \theta |$ distribution
of the soft-pions in the lower $P_T$ region, and (f) that in the higher
$P_T$ region.
The histogram
definitions are the same as those given in Figure 3.
}
\label{fig5}
\end{figure}

One of the exciting possibilities to explain this excess
is in terms of $\tilde{t}$ pair production.
A light $\tilde{t}$ at the TRISTAN energy region is still possible
if the mass difference between $\tilde{t}$ and
$\tilde{\gamma}$ is small \cite{r17,r18,r19,r20}.
In a previous paper \cite{r28}, in order to explain the
high $P_T$ excess of $D^{*\pm}$, $\tilde{t}$ pair
production was introduced with $m_{\tilde{t}}$=15 GeV
and $m_{\tilde{\gamma}}$=12.7 GeV.
These predictions are shown in Figures \ref{fig3} and \ref{fig4}
by open histograms. The experimental
cross sections agree very well with these
assumptions. The value of $m_{\tilde{t}}$ determines the
cross-section excess and that of $m_{\tilde{\gamma}}$ the event shapes,
such as the thrust and missing $P_T$ distributions. These two
distributions are shown in Figures \ref{fig5} (a)-(d).
Here, we had problems in translating the number of events into cross
sections. We thus plotted the observed number of events with the theoretical
predictions.
In addition, our Monte-Carlo generator did not include any
higher order ($\alpha_s$)
effect, such as three-jet events. One should therefore be cautious
about them. The real thrust predictions may be softer and that
of the missing $P_T$ distributions may be harder
due to the presence of three-jet events.

{}From these figures, the experimental excesses are considered to be
high-thrust events and high missing-$P_T$ events.
In order to minimize this discrepancy
in the missing $P_T$-distribution, although
we must set a larger mass difference ($m_{\tilde{t}}-m_{\tilde{\gamma}}$),
it may conflict with the recent search for $\tilde{t}$ pair
production by the VENUS experiment \cite{r32}.

If the excess is due to $\tilde{t}$ pair production, the
angular distribution with respect to the beam axis must be
different from that of the two-photon processes.
The results are also shown in Figures \ref{fig5} (e) and (f).
However, the $\tilde{t}$ assumption does not show good agreement.
The excess events dominated in the low-angle region.
We therefore can not strongly
conclude that the excess in events originated
from $\tilde{t}$ pair production. There is a possibility of a
new high-$P_T$ process in the two-photon interaction.
If this is true, it would be a serious background in future
linear-collider experiments.

\begin{figure}
\vspace{15cm}
\includegraphics{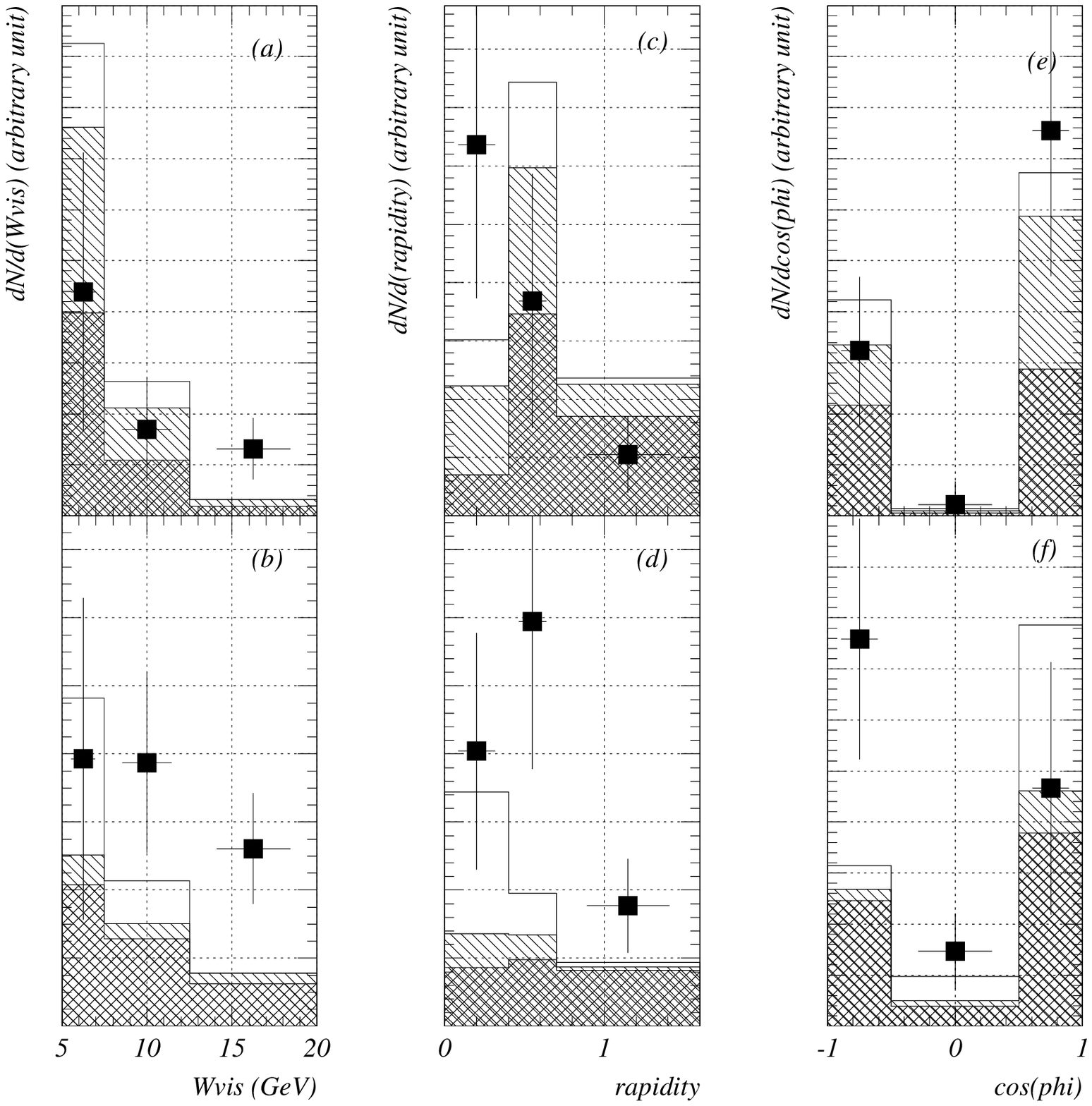}
\caption{
Various event shape distributions for the lower ($P_T<0.3$ GeV)
and higher ($P_T>0.3$ GeV)
$P_T$ events:
(a) Wvis distribution in the lower $P_T$ events, (b) that in the higher
$P_T$ region, (c) rapidity of hadronic system in the lower $P_T$ region,
(d) that in the higher $P_T$ region, (e) cosine
of the azimuthal angle between the missing $P_T$ and the soft-pion
in the lower $P_T$ region, and (f) that in the higher
$P_T$ region.
The histogram
definitions are the same as those given in Figure 3.
}
\label{fig6}
\end{figure}

In addition, we checked three more distributions, i.e., Wvis,
the rapidity of the hadronic system, and the azimuthal angle
between the missing $P_T$ and the soft-pions
as shown in Figures \ref{fig6} (a)-(f).
These distributions
does not fit to the $\tilde{t}$ assumption very well.
We definitely need more studies on this process. Especially
the electron and/or K-meson inclusive studies would be more
powerful because of higher statistics.

\subsection{Tagged events}

In a part of the data set (integrated luminosity of 90 pb$^{-1}$),
there were forward calorimeters (FCL; made of BGO) which covered the
polar angle region at between 3.2 and 13.6 degrees \cite{r21}.
We analyzed tagged events using the FCL with the same analysis.
We observed $35.5\pm12.7$ soft-pions,
where the lowest-order direct process predicts $\sim26.7$ events.
The tagged events are therefore consistent with the expectation.

\subsection{Comparison with other experimental results}


The charm measurement using the inclusive electron method
was pioneered by the VENUS group \cite{r31}. Their observed production rate
was consistent with a theoretical prediction by the
direct and resolved (LAC1) processes. However, the statistics are
especially low in the high-$P_T^c$ region. The expected $\tilde{t}$
signal rate was only a few events. We thus concluded that the VENUS
measurement was not sensitive to the $\tilde{t}$ signal, and was
consistent with our observation. A high-statistics study using high
$P_T$ electrons is awaited.


Recently, a $\tilde{t}$ search was carried out by the VENUS group
\cite{r32}. They have shown an $m_{\tilde{\gamma}}$
upper limit of $\sim 12.7$ GeV for the case when $m_{\tilde{t}}$
= 15 GeV.
This search marginally conflicts with our $\tilde{t}$
assumption.

\section{Conclusions}


We have measured the differential cross section ($d\sigma
(e^+e^-\rightarrow e^+e^- D^{*\pm} X)/dP_T$).
The $D^{*\pm}$s were identified by the transverse
momenta of soft-pions with respect to the jet-axes.
The average $\sqrt{s}$ was 58.1 GeV and the integrated luminosity
of the event sample was 198 pb$^{-1}$, respectively.
We obtained $372\pm54$ $D^{*\pm}$, which represents the
highest statistics so far obtained.
We compared the measured cross section with the theoretical
predictions.

\section*{Acknowledgement}

We appreciate discussions with Prof. M. Kobayashi (KEK),
Dr. M. Drees (Univ. of Wisconsin), M. Kr\"amer (DESY), and
J. Zunft (DESY).
We thank the TRISTAN accelerator staff
for the successful operation of TRISTAN. We also thank
all of the engineers and technicians at KEK as well as
members of the collaborating
institutions: H. Inoue, N. Kimura,
K. Shiino, M. Tanaka, K. Tsukada, N. Ujiie,
and H. Yamaoka.

\end{document}